\documentclass{sig-alternate-10pt}
\usepackage[
  pass,%
]{geometry}
\usepackage{epsfig,xspace}
\usepackage[hyphens]{url}
\usepackage{booktabs}
\usepackage{multirow}
\usepackage{crossreference}
\usepackage{array}
\usepackage{cite}
\usepackage{nth}
\usepackage{subfigure}
\usepackage[lined,linesnumbered]{algorithm2e}
\usepackage[dvipsnames]{xcolor}
\usepackage[labelfont=bf,footnotesize]{caption}
\usepackage[inline]{enumitem}

\newcommand\name{Haystack\xspace}
\renewcommand{\xref}[1]{\S\ref{#1}}

\newcommand{\custompara}[1]{\vspace{2mm}\noindent{\bf{#1}}}

\newcommand{\parax}[1]{\vspace{1.5mm} \noindent \textbf{#1:}}

\newcolumntype{L}[1]{>{\raggedright\let\newline\\\arraybackslash\hspace{0pt}}m{#1}}
\newcolumntype{C}[1]{>{\centering\let\newline\\\arraybackslash\hspace{0pt}}m{#1}}
\newcolumntype{R}[1]{>{\raggedleft\let\newline\\\arraybackslash\hspace{0pt}}m{#1}}

\providecommand{\ie}{\emph{i.e.,} }
\providecommand{\eg}{\emph{e.g.,} }

\providecommand{\is}{{\it idle\_sleep}\xspace}
\providecommand{\cn}{{\it max\_read$_{nio}$}\xspace}
\providecommand{\ct}{{\it max\_read$_{tun}$}\xspace}
\providecommand{\ic}{{\it max\_idle\_cycles}\xspace}

\providecommand{\tun}{{\tt tun}\xspace}
\providecommand{\fwder}{{Forwarder}\xspace}
\providecommand{\taslong}{{Traffic Analyzer}\xspace}
\providecommand{\tas}{{TA}\xspace}

\def\trackedflows{{394,834}\xspace}

\def\trackedapps{{761}\xspace}

\def\trackedflows{{942,836}\xspace}

\def\trackedapps{{1,340}\xspace}

\def\uniquedomains{{8,710}\xspace}

\begin{document}

\date{}

\title{\Large \bf \name: A Multi-Purpose Mobile Vantage Point in User Space}

\author{
  {\rm Abbas Razaghpanah}\\
  {\footnotesize Stony Brook University}
\and
  {\rm Narseo Vallina-Rodriguez}\\
  {\footnotesize ICSI}
\and
  {\rm Srikanth Sundaresan}\\
  {\footnotesize ICSI}
\and
  {\rm Christian Kreibich}\\
  {\footnotesize ICSI / Lastline}
\and
  {\rm Phillipa Gill}\\
  {\footnotesize Stony Brook University}
\and
  {\rm Mark Allman}\\
  {\footnotesize ICSI}
\and
  {\rm Vern Paxson}\\
  {\footnotesize ICSI / UC Berkeley}
} %

\maketitle

\thispagestyle{empty}

\begin{sloppypar}

\noindent \textbf{Abstract}
Despite our growing reliance on mobile phones for a wide range of
daily tasks, their operation remains largely opaque.  A number of
previous studies have addressed elements of this problem in a partial fashion,
trading off analytic comprehensiveness and deployment scale.  We overcome 
the barriers to large-scale deployment (\eg requiring rooted devices) 
and comprehensiveness of previous efforts by taking a novel approach 
that leverages the VPN API on mobile devices to design \name, an in-situ mobile measurement 
platform that operates exclusively on the device, providing full access
to the device's network traffic and local context without 
requiring root access.  We present the design of \name and its 
implementation in an Android app that we deploy
via standard distribution channels. Using data collected from 450
users of the app, we exemplify the advantages of 
\name over the state of the art and demonstrate its %
seamless experience
even under demanding conditions. We also demonstrate its utility 
to users and researchers in characterizing mobile traffic and 
privacy risks.

\section{Introduction}
\label{sec:intro}
Mobile phones have become indispensable aids to everyday
life by offering users capabilities that rival those of
general purpose computers.  However, these systems
remain notoriously opaque, as mobile operating systems tightly
control access to system resources.  While this tight control is useful 
in preventing unwanted application activity, it also imposes hurdles for
understanding the behavior of mobile devices, 
especially their network activity and performance.

Despite these challenges, the research community has made steady progress
in understanding mobile apps and mobile traffic over the past few years,
by using two broad classes of techniques.
One class is lab-oriented and uses
static and dynamic analysis of app source code~\cite{seneviratne2015early,pios}, 
controlled execution of apps~\cite{fahl2012eve,droids} and 
dynamic analysis~\cite{Zhang:2013:VUB:2508859.2516689}, even modifying 
the OS kernel to track app behavior~\cite{enck2010taintdroid}. 
A contrasting approach 
leverages network traces obtained from ISPs~\cite{Narseo:IMC2012,Gill:2013:BPF:2504730.2504768} or VPN tunnels that forward
user traffic~\cite{meddle} to servers in the cloud for observation.
However, each of these previous approaches faces a trade-off:

\begin{itemize}[leftmargin=*]
\item Approaches based on static and dynamic analysis do not offer access to real-world 
data. Thus far, studies that have used these approaches have 
been constrained to analysis of source code, which is not always
available, or artificial/controlled user inputs which require significant human
effort to train the techniques, contextualize the results, and minimize
false positives. One exception
is Taintdroid~\cite{enck2010taintdroid}, a modified Android version
that can analyze app behavior in real-world settings.  
However, this technique relies
on operating system modifications, which incur a significant engineering effort
to catch up with new OS releases, and forces participating users to 
install a new firmware on their devices~\cite{wijesekera2015android}.
Consequently, the scale of analysis and app
coverage they can achieve in the wild remains limited to tens of users.
\item Approaches that leverage network traffic obtain visibility into
real user behavior, at the cost of the richness of context that device-centric
approaches can obtain. For example, while this approach can capture
and analyze mobile network data, heuristics must infer which
applications generated individual flows, and detouring traffic through
third-party middleboxes complicates high-fidelity performance measurements
due to the necessarily skewed vantage point.

\end{itemize}

In this study, we present \name, the first on-device mobile measurements platform  
that is able to passively monitor app behavior and network traffic under regular 
usage and network conditions, {\em without requiring users to root the phone}. 
The latter gives \name the potential for better scalability in deployment;
users can simply install the app from 
Google's Play Store or similar markets.  This provides us the opportunity to monitor organic 
mobile network activity as generated by real users in real networks using real 
mobile apps---all from the vantage point of the device. 
This combination of ease of deployability and the high-fidelity vantage point allows \name to hit
a sweet spot in the trade-off between 
scalability and richness of data. 

Similar to previous approaches~\cite{meddle}, \name leverages Android's standard
VPN interface to capture outbound packets from applications. However,
rather than tunneling the packets to a remote VPN server for 
inspection, \name intercepts, inspects, and forwards the user's traffic to its intended
destination. 
This approach gives us raw packet-level access to outbound packets as well as flow-level 
access to incoming traffic without modifying the network path, and without
requiring permissions beyond those needed by the VPN interface.
\name therefore has the ability to monitor network activity in the proper 
context by operating locally on the device. For example, a TCP connection 
can be associated with a specific DNS lookup and both can be coupled with the 
originating application. Further, we design \name to be extensible with 
new analyses and measurements added over time (\eg by
adding new protocol parsers and by supporting advanced measurement
methods such as reactive measurements~\cite{allman2008reactive}), 
and new features to attract and educate users (\eg ad block, malware detection, privacy
leak prevention and network troubleshooting).

\name is publicly available for anyone to install
on Google Play and has been installed by 450 users to date~\cite{GplayListing}. 
We discuss the design and implementation of \name in
\xref{sec:systemdescription} and \xref{sec:implementation},
and evaluate its performance and resource use in
\xref{sec:evaluation}. Our tests
show that \name delivers sufficient throughput (26--55~Mbps) at low latency
overhead (2--3~ms) to drive high-performance and delay-sensitive applications
such as HD video streaming and VoIP without noticeable performance degradation
for the user.

While we consider our \name implementation prototypical in some
respects (such as UI usability for nontechnical users), it has already
provided interesting insights into app usage in the wild:
in~\xref{sec:applications} we present preliminary findings about the adoption of encryption
techniques, report on local-network traffic interacting with IoT
devices, study app provenance and the use of third-party tracker
services, and give an outlook on potential future applications.

\section{Related Work}\label{sec:background}

\begin{table*}[t!]
\scriptsize
\centering
\begin{tabular}{|L{2.2cm}|C{2cm}|C{1.55cm}|C{2.9cm}|C{1.55cm}|C{2cm}|C{2.8cm}|}
\hline
{\bf Approach}   & {\bf Scale} & {\bf Real-world operation} & {\bf Comprehensiveness} & {\bf Local Operation} & {\bf App coverage} & {\bf OS compatibility} \\
\hline
ISP traces & Large-scale & \checkmark &  & & All apps & All versions/platforms \\
Remote VPN & Crowdsourcing & \checkmark &  & & Crowdsourcing & All versions/platforms \\
Static analysis & Resource-bound & & \checkmark & \checkmark & $\sim 1000$ apps & Limited \\
Dynamic analysis & Resource-bound & & \checkmark & \checkmark & $\sim 100$ apps & Limited \\

\hline
\end{tabular}
\vspace{-1mm}
\caption{Comparison between different measurement approaches in the mobile environments. It should be noted that these aspects/features are not easily comparable in a binary manner and the comparison provided here is merely qualitative.}
\label{table:method_comparison}
\end{table*}

Previous studies have leveraged a variety of techniques for
understanding privacy risks of mobile apps 
and their behavior in the network.  As noted earlier,
each approach made trade-offs between having access to real 
user behavior and device context.   
We classify the prior work into the following four categories.

\custompara{Dynamic app analysis: } This approach calls for
running an app in a  controlled environment such as a virtual
machine~\cite{Zhang:2013:VUB:2508859.2516689} 
or an instrumented OS~\cite{enck2010taintdroid,droids}.  The app is then
 monitored as it conducts its pre-defined set of tasks, with the results
indicating precisely how the app and system behave during the test
(\eg whether the app exfiltrated data).  While this
approach provides useful insights, the 
workload (which does not represent real-world operation) and difficulty of
deploying custom firmware on users' phones (sacrificing scale)
means that the results do not directly speak to normal users' activity. 
To overcome the lack of 
user input, studies that rely on dynamic analysis require  
``UI monkeys''~\cite{wong2016intellidroid, anand2012automated} to generate synthetic 
user-actions.

\custompara{Static app analysis: } This technique involves analysis of the app code,
obtained by decompiling app binaries, via symbolic
execution~\cite{Yang:2013:AAS:2541806.2516676}, analysis of control flow
graphs~\cite{pios,pscout}, by auditing third-party library 
use~\cite{seneviratne2015trackingpaidapps,chen2014information}, 
through inspection of the Android permissions 
and their associated system calls~\cite{leontiadis2012don,pscout}, 
and analysis of app properties (\eg whether apps employ secure communications)
~\cite{fahl2012eve,Georgiev:2012:MDC:2382196.2382204}.  While static analysis typically
provides good scale with analysis of over 10K apps in several studies
~\cite{fahl2012eve}, (modulo computational resources)  
this strategy does not reflect the behavior
of apps in the wild, and typically requires a good amount of manual
inspection.  Furthermore, the analysis may under- or
over-state the importance of certain code paths since it lacks a
notion of how users interact with the apps in practice.

\custompara{Passive traffic analysis:} A number of
studies rely on volunteers with rooted phones that allow their traffic to get 
recorded by {\tt tcpdump}~\cite{FalakiFirstLook,Qian:2011:PRU:1999995.2000026,
Huang:2012:STC}  or {\tt iptables}~\cite{vallina2013rilanalyzer,aucinas2013staying}. 
These methods are challenging to deploy at scale.
To obtain larger-scale data, %
other projects study the behavior of 
mobile devices by observing their network traffic either at a large ISP with millions of
users~\cite{Gill:2013:BPF:2504730.2504768,Narseo:IMC2012,Shafiq:2013:FLC:2465529.2465754}
or by forwarding traffic through a remote VPN
proxy that also modifies the network path~\cite{meddle,antmonitor,
2015arXiv150406093V}. 
As a result, these studies contain a large variety of apps and mobile platforms 
but they lack device context for accountability
and accuracy (\eg mapping flows to originating apps). 
While this can be alleviated by pairing a remote VPN proxy with client-side 
software to provide context to the remote VPN server~\cite{antmonitor}, the solution
still alters the network path 
by rerouting traffic to the VPN server, hence providing an
unrealistic view of the performance aspects of real mobile traffic.
PrivacyGuard~\cite{Song:2015:PVP:2808117.2808120} uses a technique similar 
to the one used by \name to intercept user traffic to detect simple 
instances of private information leaks (\eg device ID and location), 
but it does not aim to offer the depth and versatility offered by \name
as a measurement platform.

\custompara{Active mobile network measurements:} Google Play (and, on
a smaller scale, the Apple Store) contains a number of tools for
active mobile network measurements.  Examples include
Ookla's SpeedTest~\cite{ooklaspeedtest}, the
FCC Speed Test~\cite{fccspeedtest}, network scanners to build network coverage 
maps~\cite{opensignal}, and
comprehensive measurement tools
such as My Speed Test~\cite{myspeedtest}, Netalyzr for Android~\cite{netalyzrGPlay},
NameHelp~\cite{namehelp}, and MobiPerf~\cite{mobiperf}.
Such tools provide valuable insight into 
network performance~\cite{sommers2012cell,nikravesh2015mobilyzer} 
and operational aspects of ISPs  
such as middlebox deployment~\cite{vallina2015beyond}
and traffic discrimination~\cite{kakhki2015identifying}. However, 
despite the fact that 
active measurement techniques typically provide an accurate snapshot of actual network conditions, they do not study network performance of installed apps in real-world situations.

Table~\ref{table:method_comparison} provides a high-level comparison of
each of the measurement approaches. 
As we can see, none can simultaneously observe real-world operation while providing  
comprehensive data at scale. This trade-off has prevented 
the research community from 
exploring in detail many aspects of the mobile ecosystem. We will revisit the 
comparison between \name and state of the art techniques in~\xref{sec:applications}, 
after we present its design, implementation, and evaluation.

\section{Haystack Overview}
\label{sec:systemdescription}

Our goal with \name is to help researchers avoid the trade off between accessing
device context and 
the ability to measure real-world phone usage at scale. 
The crux of \name is its ability to observe network communication \emph{on
the mobile device itself}.  Since 2011 (version~4.0), Android
has provided a VPN API that enables developers to create a virtual \tun
interface and direct {\em all} network traffic on the phone to the interface's
user-space process.  To enable this functionality, the client app
requests the {\tt BIND\_VPN\_SERVICE} permission from the user,
which, crucially, does not require a rooted device.  The API typically
drives VPN client applications that forward traffic to a remote
VPN server~\cite{vpnservicedoc}.  Instead of relaying
packets to a remote VPN server, \name performs two
high-level operations in parallel.  First, it sends a copy of the
bidirectional packet stream to a background process that analyzes the
traffic off-path. Second, it uses the packet headers to maintain user-space
network sockets to remote hosts and relays %
data via
these sockets.

\name is available in the Google Play Store~\cite{GplayListing} and has been downloaded a 
total of 450 times. 
A number of apps in Google Play leverage the {\tt BIND\_VPN\_SERVICE}
permission for non-VPN tasks.  While we are not aware of any apps
taking traffic processing to the level we realize in \name, 
tPacketCapture~\cite{tpacketcapture} and SSL Packet
Capture~\cite{sslpacketcapture} take advantage of the VPN API to
record approximate packet traces via a user-space application. As with \name,
these traces are approximate since the app does not have access to raw packets 
via Java's socket interface. 
A related app, NoRoot Firewall~\cite{norootfirewall} allows  
mobile users to block traffic generated by specific
apps and generate connection logs. 

\subsection{Ethical Considerations}

\name's ability to observe real-world user data raises many ethical 
considerations~\cite{ethics}. We leverage the fact that \name runs 
 on the user's device to do the bulk of processing on the device and only send back summary statistics (\eg domains contacted and protocols used) and 
by under no-circumstances user's raw traffic. 
We aim to minimize the amount of data sent back while maximizing it's utility.
In consultation with the IRB at UC Berkeley, we developed a protocol that 
strikes a balance 
and only collects data needed for the studies at hand without uploading
any personal information. This precludes certain types of detailed or longitudinal studies, 
which may be possible with future coordination with the IRB. 

Additionally, we implement informed consent and opt-in in \name.
First,  \name must be explicitly installed by
the user and granted permission to observe traffic.  Second, we
require users to opt-in a second time before we analyze encrypted
traffic as described in \xref{sec:tls}. TLS interception is explained to 
the user in detail before they are given the option to install the CA 
certificate needed to intercept encrypted traffic. %
If the user chooses not to install the CA certificate, the TLS interception 
module is disabled until they explicitly choose to install the CA certificate and 
manually enable TLS interception in the settings. 
Our opt-in process aims 
to make the data collection as transparent as possible and provide users control over the process. 
While our IRB has reviewed our current 
approach and has deemed our work as not involving human subjects
research, we maintain an active dialogue and we will seek their feedback before collecting any additional piece of information.

\begin{figure}[t]  \centering
  \includegraphics[width=1\columnwidth]{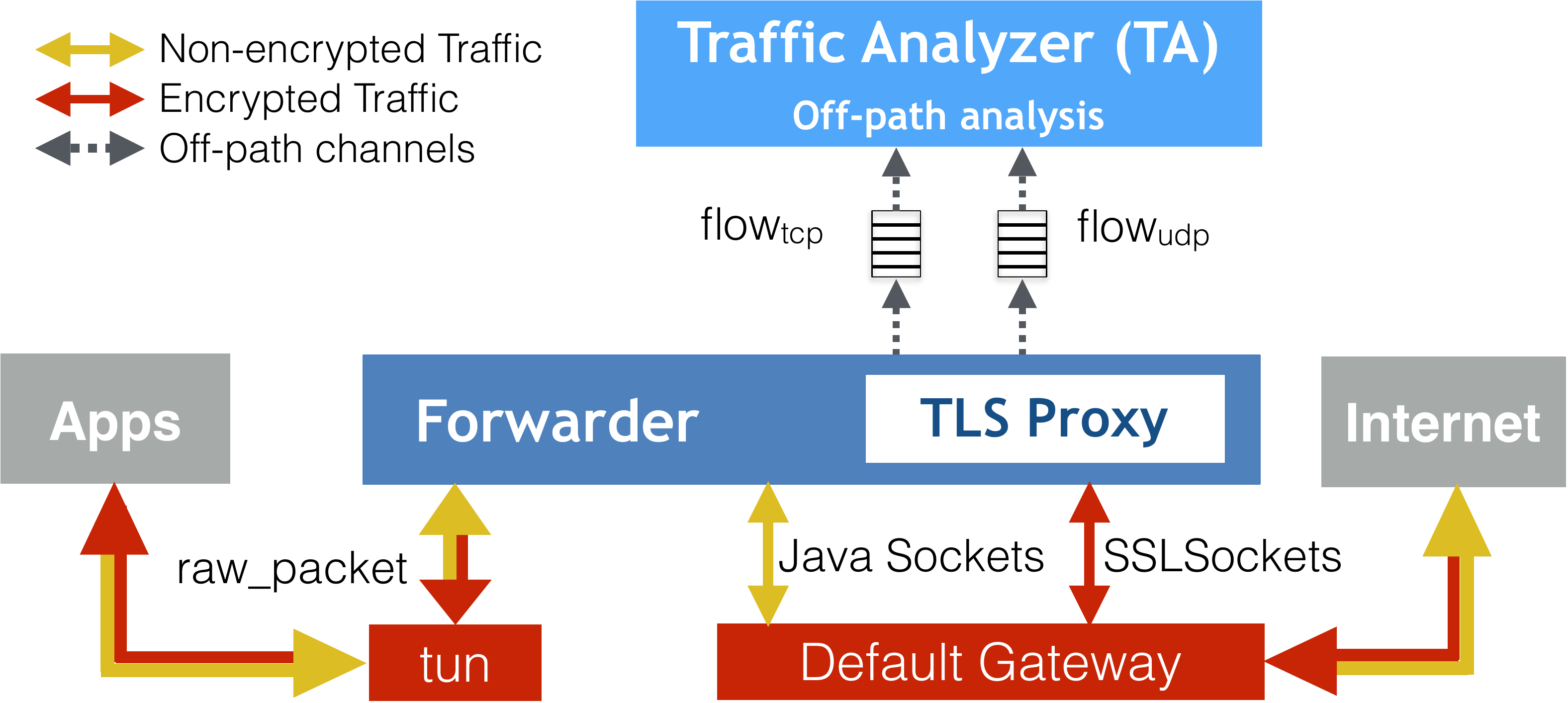} 
  \caption{The \name architecture, highlighting
   system components and data forwarding channels. 
   Solid lines represent the actual forwarding path for traffic generated by 
   mobile apps even if encrypted (which is handled by our optional TLS proxy), 
   while dashed lines represent the off-line path used for privacy and 
   performance analysis.}
  \label{fig:hs_arch}
\end{figure}

\begin{figure}[t]  \centering
  \includegraphics[width=1\columnwidth]{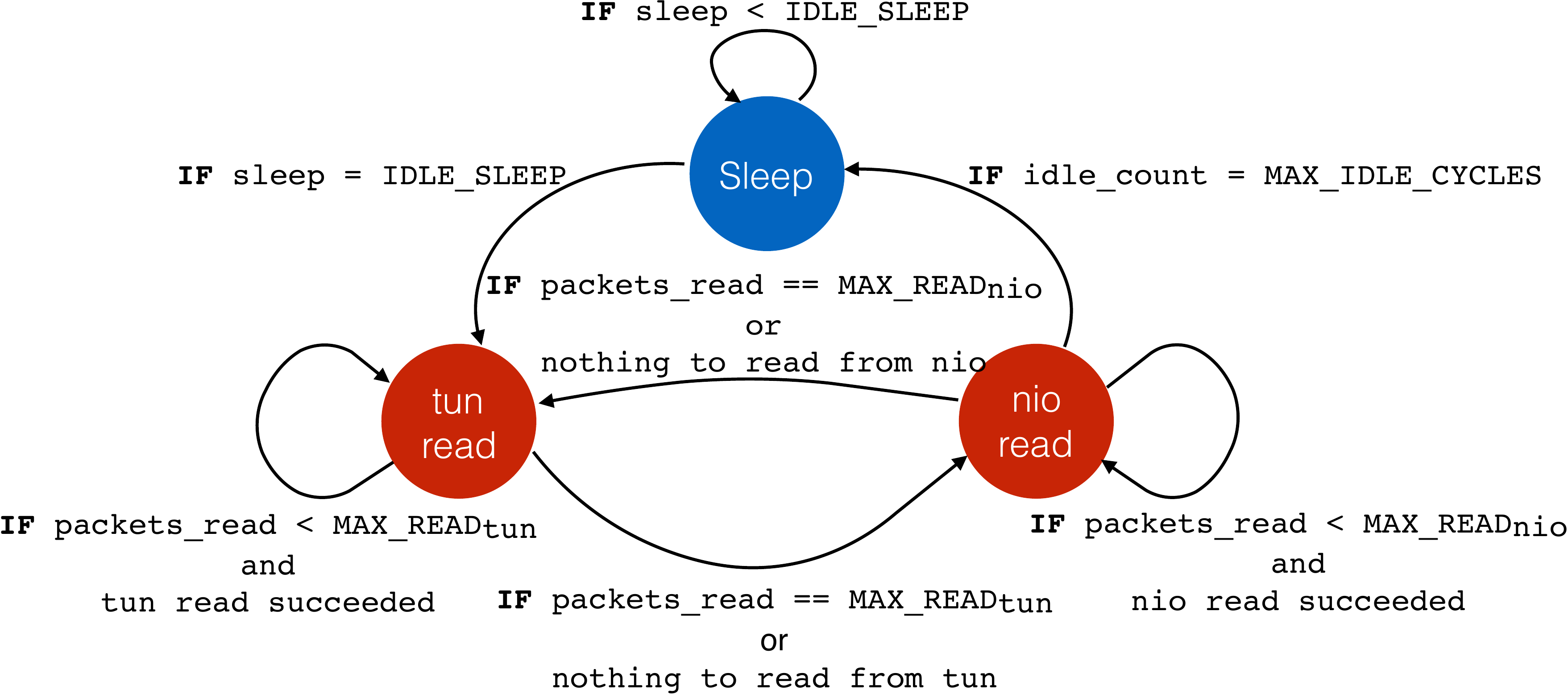} 
  \caption{\name's \fwder state machine. It controls read/write
  operations and transitions between \tun interface, Java NIO socket, 
  and sleep states. 
  The idle count variable increments when both \tun and NIO do not
  succeed, \ie there is nothing to read. Each read operation from the \tun 
  interface potentially becomes a write
  operation for a NIO socket and vice versa.  
}
  \label{fig:state_machine}
\end{figure}

\section{System Design}
\label{sec:implementation}

To intercept and analyze traffic on resource-constrained
devices in user space, we must address several design challenges. A
key issue is that the \tun interface exposes raw IP packets to
\name.  A natural way to deal with these would be to shuttle a copy
to our analysis engine and then drop the packet on the network via a
raw socket. 
However, non-privileged apps do not have access to raw
sockets and therefore we must rely on regular Java sockets to
communicate with remote entities.  This means that, as opposed
to transparent L3 and L4 proxies that operate at a single 
layer of the protocol stack on both sides (with root
privileges), \name has to
bridge packet-level communication on the host (\tun) side and
flow-level interaction with the network side. Operating in
mobile phones in user space requires
careful design considerations to minimize \name's  
impact on device resources, battery life, app performance and user experience. 
Figure~\ref{fig:hs_arch} illustrates the \name architecture, which
includes two major components, the {\em \fwder} and the {\em
  \taslong} (TA).

\subsection{The \fwder}\label{subsec:forwarder}

The \fwder performs two key functions: 
($i$) it performs transparent bridging between packets on the \tun
interface and payload data on the regular socket interface and
($ii$) it forwards traffic to the \tas for analysis.

\subsubsection{Flow reassembly}\label{sec:handlingstate}

The \fwder receives raw IP packets from \tun.
The \fwder therefore acts like a
layer 3/layer 4 network stack: it extracts the payload from the raw
packet and sends it to its intended destination through a regular
Java socket (implemented 
using non-blocking NIO sockets~\cite{javanio}).   
To accomplish this, the \fwder extracts flow state
from the packet headers (IP, as well as UDP or TCP) for packets
arriving on the \tun interface and maps it to a given Java socket
(it creates new sockets for new flows arriving on the \tun interface).
It also maintains this state so that
it can marshal data arriving from remote hosts on the sockets 
back into packets for transmission to the app via the
\tun interface. \name has dual-stack support and 
its routing tables correctly forward DNS and IPv6 traffic 
through the \tun interface to prevent traffic 
leak~\cite{perta2015glance}.

\custompara{Handling UDP and TCP: } The \fwder needs to
maintain state for UDP and TCP flows. A simple flow-to-socket
mapping suffices for connectionless UDP, since header
reconstruction remains straightforward. Since TCP provides connection-oriented and
reliable transport, we need to track the TCP state machine and
maintain sequence and acknowledgment numbers for each TCP flow.
We segment the data stream received from the socket and synthesize TCP
headers to be able to forward the resulting packets to the \tun
interface for delivery to the app.
When we read a SYN packet from the \tun
interface, we create a new socket, connect to the target and
instantiate state in \name.  After the OS establishes the socket we
return a SYN/ACK via the \tun interface. We similarly relay connection
termination. We discuss \name's lack of support for 
non-TCP/UDP traffic in~\xref{sec:implementation:limitations}.

\custompara{Efficient packet forwarding:} 
The \fwder must balance application and traffic performance with power and CPU usage
on the device.  This task is challenging because the \tun
interface does not expose an event-based API.  We therefore
implement a polling scheme that periodically checks both the \tun
interface and Java sockets for arriving data.

Figure~\ref{fig:state_machine} shows the state machine of the \fwder.
It reads up to \ct packets  
from the \tun interface or up to \cn packets from the 
socket (NIO)\footnote{Despite the inability to count
    packets from socket read/write operations, 
    we count the number of packets 
    generated and sent back through the \tun interface.
}
interface before switching to the other interface, hence preventing
either operation from starving. The \fwder immediately transitions to the
other read state if it cannot read data in the current state.
Each read from the \tun interface potentially becomes a write
operation for a socket and vice versa, the exception being pure
TCP ACKs from the \tun interface. We discard these, as their
effect gets abstracted by the socket interface and therefore they do not
require forwarding.  Writes to the \tun interface complete quickly
and socket writes do not block, so we perform writes as soon as we
have data to send.  If it cannot read data from either interface for
\ic consecutive iterations, the \fwder will sleep for \is~ms.
While this strategy reduces power consumption during idle periods,
it also imposes higher latency on packets that arrive during these
idle periods when polling happens at coarse intervals.  We consider
the tradeoff between resource conservation and performance in depth
in \xref{sec:cpubatteryeval}.

\subsubsection{TLS Interception}\label{sec:tls}

Many mobile applications have adopted TLS as the default
cryptographic protocol for data communications.  This is a
double-edged sword, as it helps protect the integrity and privacy of
users' transactions but also allows apps to conceal their network
activity.  With the user's consent, \name
employs a transparent man-in-the-middle (MITM) proxy for TLS
traffic~\cite{sslmitm}.  At install time \name requests the user
allow the installation of a self-signed \name CA certificate in the
user CA certificate store.  We customize the message shown to users
at this time to explain why \name intercepts encrypted traffic.

Once equipped with a certificate, the \fwder monitors TCP streams
beginning with a TLS ``Client Hello'' message and forwards these
flows---along with flow-level meta-information the proxy requires in
order to connect to the server (\eg IP address, port, SNI)---to the
TLS proxy.  The proxy uses this information to connect to the remote
host and reports back to the \fwder whether the connection was
successful.  After successfully establishing a connection to
the remote host, the proxy decrypts traffic arriving on one
interface (\tun or socket) and re-encrypts it for relay to the other while providing a clear-text version to the
\tas for analysis.

\custompara{Dealing with failed TLS connections:} 
As in any commercial TLS proxy, \name will
be unable to proxy flows when the client application ($i$) uses TLS
extensions not supported by \name~\cite{blake2006transport},%
\footnote{Currently, \name only supports the SNI extension.}
($ii$) bundles its own trust store, or ($iii$) implements
certificate pinning.  Likewise, failure occurs when the server
expects to see certain TLS extensions not supported by \name in the
``Client Hello'' message or performs certificate-based client
authentication.  We add connections with failed TLS handshakes to a
whitelist that bypasses the TLS proxy for a period of five minutes.
Experience with our initial set of users indicates that apps recover
gracefully from TLS failures.
After five minutes we remove the app
from the whitelist to account for transient failures in the
handshake process. While we cannot decrypt such flows, we can still
record which apps take these security measures and potentially
communicate more securely for further analysis.

\custompara{Security considerations:}
Android provides support for third-party root certificate installation.    
This is a feature required by enterprise networks to 
perform legitimate TLS interception.  For increased security, \name
generates a unique certificate and key-pair for each new
installation of the app.  Additionally, \name saves the private key
to its private storage to prevent other applications from accessing
it. While these precautions still permit malicious applications with
root access to retrieve the key, such apps can already tap into the
user's encrypted traffic without using \name's CA certificate (\eg
by surreptitiously injecting their own CA certificate into the
system's trust store).

\subsection{\taslong}
\label{sec:implementation:is}

The \taslong (\tas) processes flow data captured by the \fwder. The
\tas operates in near real-time but off-path, \ie outside the
forwarding path of network traffic. The \tas augments flows with
contextual information gathered from the OS for further analysis.
The analyses are protocol-agnostic, and \tas supports protocol parsers to 
parse flow contents before they are analyzed. We currently support 
TLS, HTTP, and DNS protocol parsers to analyze the traffic and extract 
relevant information, decompressing and decoding compressed and encoded data before they are searched by the DPI module for private information leakages. New protocol parsers can be added to \tas in case we see new protocols getting widely adopted.

\custompara{Why off-path analysis?}  The \tas could potentially
negatively affect the user experience if done in the forwarding
path.  Analysis of network traffic can range from simple (\eg tracking 
 packet statistics) to quite complex (\eg parsing protocol
content) and therefore can consume 
valuable CPU cycles and if conducted as part of traffic forwarding
could increase latency. However, as we will discuss in~\xref{sec:applications}, 
certain aspects of mobile apps and networks must be measured in-path 
as in the case of traffic performance analysis. 

\custompara{Secure and efficient IPC in Android:} 
Unfortunately, low-latency communication between Android services can prove tricky to realize,
especially in multi-threaded systems. In our implementation we use
Java's thread-safe queues for communication between the Forwarder and \tas modules. This allows the modules 
to communicate without exposing their data to other (malicious) 
apps as would be the case if the file system 
or localhost sockets were used~\cite{socketsIPCAndroid}. 
In~\xref{sec:intelserveval} we evaluate the overhead of using thread-safe 
queues to enable communication between the \fwder and \tas.

\custompara{Application and entity mapping:} One of the basic
functions the \tas provides is to map TCP and UDP flows to the
corresponding apps.  We do this via a two-step process:
($i$) extract the PID of the process that generated the flow from
the system's {\tt proc} directory, ($ii$) map the PID to an
app name using Android's Package Manager API.  Compared to
network-based studies which rely on inferences---\eg using the HTTP
User-Agent or destination IP address---to couple apps and
flows~\cite{meddle,recon,Narseo:IMC2012}, our approach allows
highly accurate flow-to-app mappings.  Since reading the PID and
mapping applications requires file-system access, we cache recently
read results to minimize overhead.

The \tas also provides the ability to analyze protocols in depth.
For example, the \tas tracks DNS
transactions to extract names associated with IP addresses, allowing
us to map flows to target domains rather than just IP
addresses. This is especially important for non-HTTP flows (\eg
QUIC, HTTPS) where the hostname may not be readily available in
application layer headers.  Mapping IPs to their hostnames gives us
the opportunity to distinguish apps sending data to their own
backend as opposed to third-party ad/analytics services or CDNs,
even if both reside in the same cloud service
provider~\cite{bermudez2012dns}.  Further, the \tas can perform
traffic characterization based on domain, without analyzing
application-layer headers (\eg HTTP {\tt Host} header).
We demonstrate how these capabilities in the \tas can enable studies
like per-app protocol usage and user tracking detection in
\xref{sec:applications}.

\subsection{Limitations and other considerations}
\label{sec:implementation:limitations}

\custompara{Protocol support:} Android limits us to only TCP and UDP
sockets via Java's APIs, thus excluding protocols such as ICMP.
As of today, this limitation only seems to 
affect a small number of network troubleshooting tools.
The \fwder provides IPv6 support, except for extended headers. We
have not noticed any  issues for IPv6 flows due to this
limitation.

\custompara{Recovery from loss of connectivity:} The VPN service (and therefore the
\fwder) gets disrupted when users roam between different networks
such as 3G and WiFi or different WiFi networks,
or when a network disconnection occurs.  \name identifies such events
and attempts to reconnect seamlessly.  Similarly, phone calls
disable all data network interfaces, thus stopping the VPN service.
While currently this disables \name, we are working on using Android
APIs to identify when the calls complete to transparently
restart the VPN.

\custompara{Vendor-custom firmware:}
Many device vendors block and interfere with standard
Android APIs. One case is  Samsung's KNOX SDK---only available for Samsung 
licensed partners---which prevents third-party VPN applications from 
creating virtual interfaces~\cite{knoxapi}. 
Likewise, some vendor-locked firmwares also prevent \name from
intercepting TLS traffic by blocking CA certificate installation. We have thus far primarily encountered this issue on Samsung phones.

\custompara{DPI and arms race:} Malicious agents will always have an incentive
to not being identified. Against our best efforts to parse and extract 
information from popular protocols, inflate compressed streams, and intercept 
conventional TLS-encrypted flows; as well as \name's ability to support newer 
protocol (\eg QUIC and new TLS extensions) 
as mobile apps and the mobile ecosystem as a whole evolve, 
some apps will still be able to exfiltrate private information through obfuscation 
and encryption schemes that are not supported by \name. Since \name would fall 
short of studying these instances, we acknowledge that there is a possibility 
of an arms race between privacy-invasive and malicious apps and approaches like 
\name. 

\section{Performance Evaluation}\label{sec:evaluation}

We have implemented \name as a user-level Android app per the design
given above.  Our implementation leverages a number of external
libraries for tasks such as efficient packet parsing~\cite{jpcap},
 IP geo-location~\cite{maxmind}, 
data presentation~\cite{MPAndroidChart}, 
and TLS interception~\cite{sslmitm}.  The \name
codebase---excluding the external libraries and XML GUI
layouts---spans 15,000 lines of code.  In this section we evaluate
to what extent we achieve our goal of real-time monitoring without
burdening the device's resources in practice and under stress
conditions.

\subsection{Testbed and Measurement Apparatus}

To evaluate \name performance in a controlled setting, we set up a testbed with
a Nexus 5 phone connected to a dedicated wireless access point over
a 5~GHz 802.11n link.  We also connected a small server to the
access point via a gigabit Ethernet link. We minimize background
traffic on the phone by only including the minimal set of
pre-installed apps and not signing into Google Services.  We measure
the latency of \name using simple UDP and TCP echo packets.  For
non-TLS throughput tests, we use a custom-built speed-test that
opens three parallel TCP connections to the server for 15 seconds in
order to saturate the link. We test uplink and downlink separately.
For profiling TLS establishment latency and downlink speed-test, we
cannot use our speed-test, as it does not employ a TLS
session. Instead, we download 1~B and 20~MB objects over HTTPS from
an Apache v2 web server with a self-signed x.509 certificate. We
repeat each test 25 times.

While our testbed allows us to explore many parameters within the
design space, Android's VPN security model precludes full automation
of our experiments as it requires user interaction to enable/disable the 
\tun interface. We focus on the impact of \ic
and \is and fix \ct and \cn to 100~packets which favors downlink
traffic.

\subsection{CPU and Power Overhead}
\label{sec:cpubatteryeval}

CPU usage impacts interactivity of foreground apps and as a result, 
the user experience. 
We therefore investigate the impact of how idle a device must be before
starting periodic polling (for a maximum of \ic~cycles) and how often we poll for new
traffic after a device is deemed idle (\is~ms) on CPU load and
battery life. 

\begin{figure}[t]  
    \centering
    \includegraphics[width=1.0\columnwidth]{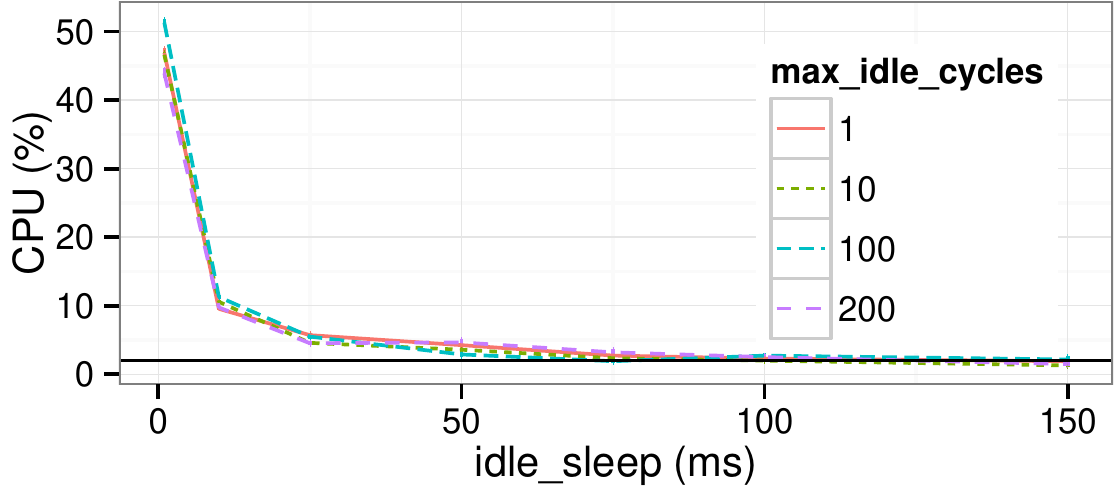}
    \caption{\name's CPU overhead for different \ic and \is
        configurations. 
	The horizontal line indicates
	the aggregated average CPU load of all apps running on the 
	background for reference. }    
    \label{fig:cpu_cost}
\end{figure}

\custompara{CPU load:} 
Mobile phones remain idle most of the time~\cite{aucinas2013staying,vallina2012energy}. 
As a result, optimizing \name's performance in this scenario is 
essential to minimize its impact on limited system resources, in particular
on battery life.  
The base CPU usage of the Nexus 5 is 2\% in the
absence of \name, when the system is idle with its screen off and normal 
background activity from installed apps. 
Figure~\ref{fig:cpu_cost} shows the impact of \ic and \is on CPU usage when 
enabling \name. We find that \is has 
the most significant effect on CPU load, which is unsurprising as
this parameter dictates how long the app sleeps and therefore does
not consume CPU.  
With \is set to  1~ms, the CPU load varies between 45\%
and 55\% for different values of \ic with the \fwder polling the interfaces at a 
high frequency. 
CPU usage drops sharply as we increase \is, to 10.5\% and 4.6\% with \is at 10~ms 
and 25~ms, respectively. In contrast to \is, \ic shows little influence on CPU overhead,
particularly at \is values greater than 10~ms. This is because 
we measure \ic in loop cycles (cf. Figure~\ref{fig:state_machine}) which take a small
fraction of 1~ms each. For \is of 100~ms and \ic of 10 or 100~cycles
the overhead of  
\name is negligible, with the CPU usage close to the base CPU usage (horizontal 
line in Figure~\ref{fig:cpu_cost}). We consider an \is value of 100~ms
ideal for operating during idle periods (delay-tolerant) 
and an \is value of 10~ms during interactive periods. In the following subsections,
we will evaluate the impact of \is in traffic performance.

\custompara{User experience during interactive periods:} We next profile \name's overhead 
under heavy load. To do so we run \name and
stream a 1080p YouTube video. This stresses packet forwarding, 
CPU usage, and the TLS Proxy, since YouTube delivers the video over TLS.
Crucially, we do not observe delay, rebuffering events, or noticeable change in resolution during the video replay,
suggesting that \name's performance can keep up with demanding applications.

\custompara{Power consumption:}
We use the Monsoon Power Monitor~\cite{monsoon} to directly measure the power 
consumed by \name on a BLU Studio X Plus phone running Android 
5.0.2.~\footnote{We faced several instrumentation challenges
	that impeded measuring \name's power consumption on a Nexus 5.} 
We removed the battery and replaced it with the power meter set to
emulate the phone's standard 3.8V battery.  We then record the power
consumed during various situations.  Table~\ref{table:battery_stats}
summarizes the results for each scenario across 10~trials with \ic
set to 100~cycles and \is set to 1~ms. This configuration represents
the worst-case (cf. Figure~\ref{fig:cpu_cost}) as \name sleeps for
only 1~ms before polling the interfaces again.  Unfortunately, due
to hardware limitations, we could not measure \name's power
consumption with the screen off but, for that scenario,
we can use \name's CPU overhead as a proxy~\cite{vallina2012energy}. 
During idle periods with the
screen active, \name increases power consumption by 3\% (similar to
the CPU increase).  The overhead of \name increases to 9\% while
streaming a YouTube video.

\begin{table}[t!]
\centering
\scriptsize
\begin{tabular}{|L{2.55cm}|C{3.7cm}|C{1.2cm}| }
\hline
{\bf Test Case} & {\bf Power(mW) Mean/SD}  & {\bf Increase}  \\ \hline
Idle & 1,089.6~/~125.9 & \multirow{2}{*}{+3.1\%}   \\ \cline{1-2}
Idle (\name) & 1,123.8~/~150.4  &  \\ \hline
YouTube & 1,755.3~/~35.5 & \multirow{2}{*}{+9.1\%}  \\ \cline{1-2}
YouTube (\name) & 1,914.4~/~16.1 &   \\ \hline
\end{tabular}
\caption{Power consumption of \name when \ic is 100 cycles and \is is 1~ms
    in different scenarios. The percentage indicates the increase when running \name.
}
\label{table:battery_stats}
\end{table}

\begin{figure}[t]  \centering
  \includegraphics[width=1\columnwidth]{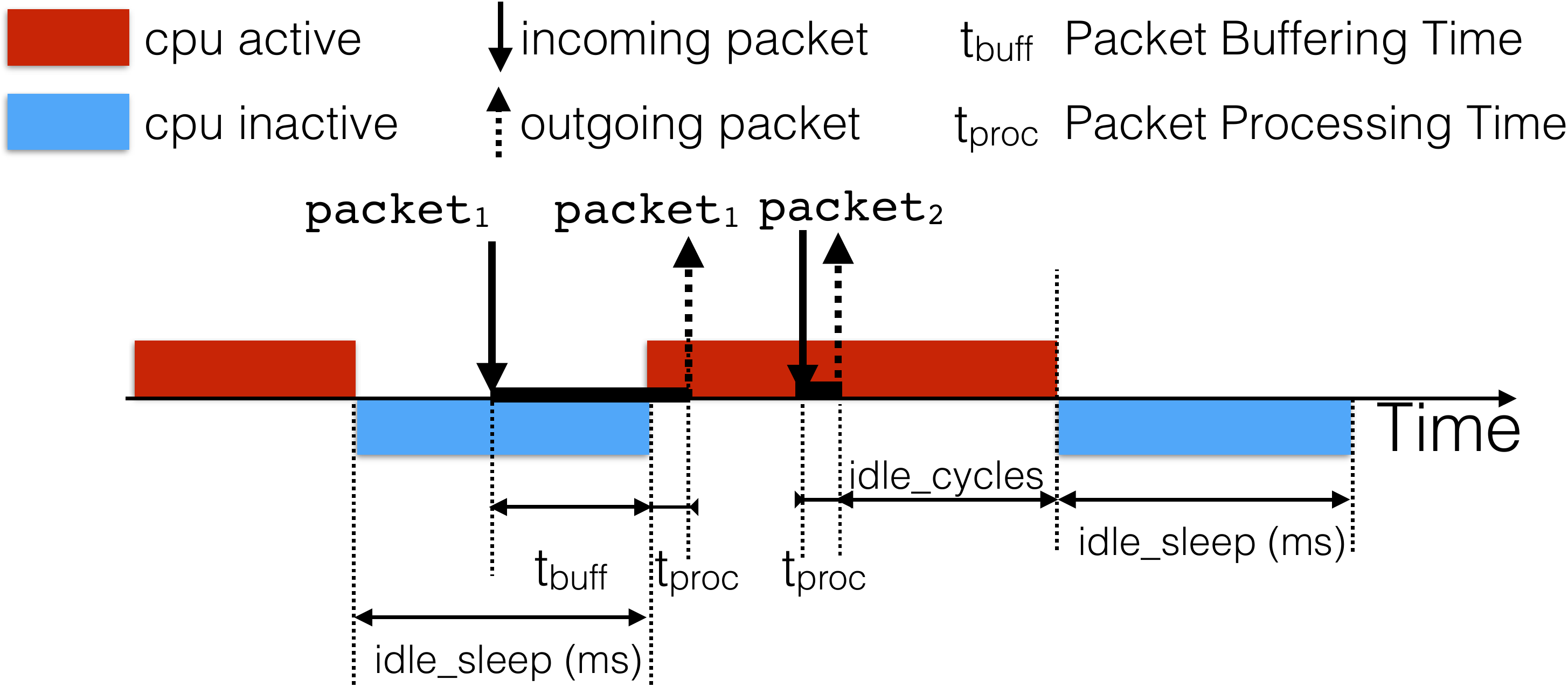} 
  \caption{Latency added by \is and \ic on
	packets arriving during periods of activity,
	or inactivity.}
  \label{fig:packet_latency}
\end{figure}

\begin{figure*}
  \subfigure[UDP latency.]{
    \includegraphics[width=0.315\linewidth]{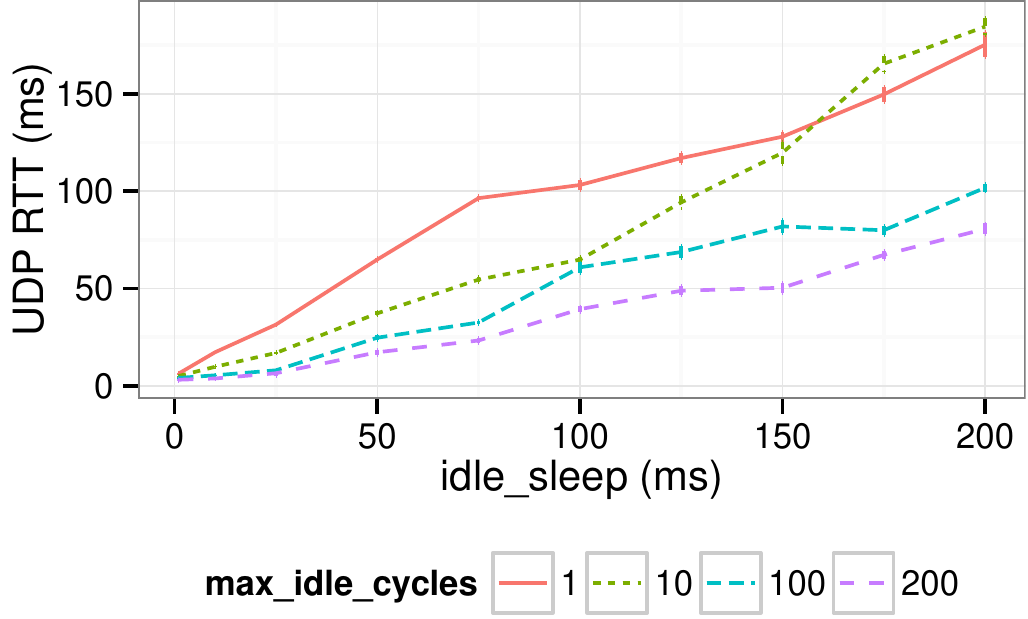}
    \label{fig:latency_overhead}
  }
  \hfill
  \subfigure[TCP connection time.]{
    \centering
    \includegraphics[width=0.315\linewidth]{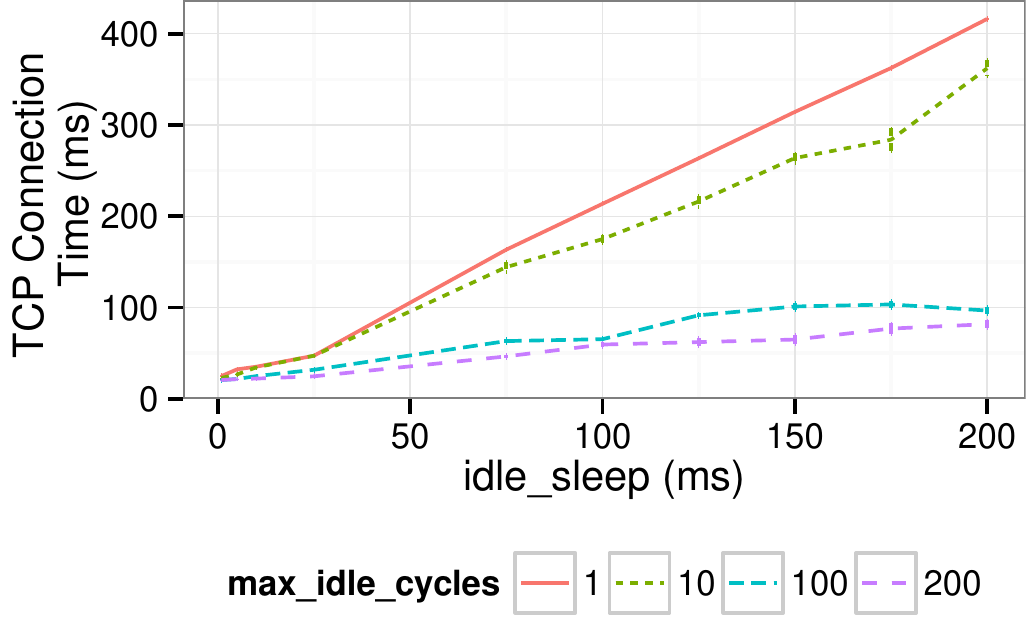}
    \label{fig:tcp_lat_overhead}
  }
  \hfill
   \subfigure[TCP Throughput.]{
      \centering
      \includegraphics[width=0.315\linewidth]{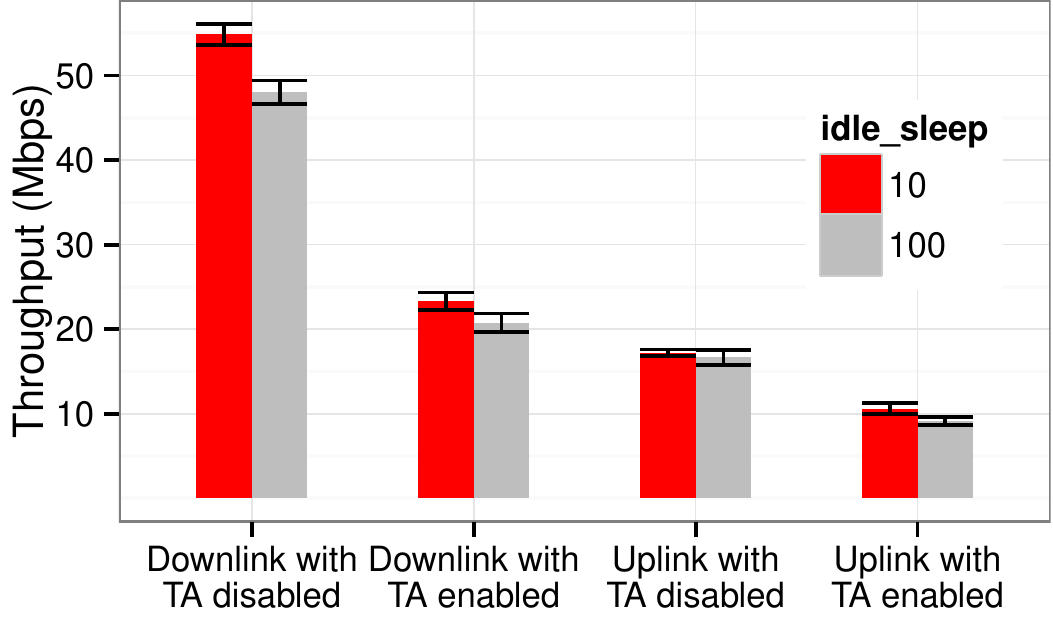}
      \label{fig:throughput_overhead}
    }
  \vspace*{-1em}
  \caption{\name performance (UDP latency, TCP connection time,
    and TCP throughput) for different \is and \ic 
    configurations. For the throughput evaluation, we fix
    \ic to 100 cycles, also showing the impact of enabling 
    the \tas. The maximum TCP throughput for this link
    is 73 and 83~Mbps uplink and downlink, respectively.}
  \label{fig:perf_eval}

\end{figure*}

\subsection{Latency Overhead}

\name suspends polling during periods of inactivity to conserve battery. However, suspending polling also
increases latency for packets that arrive during loop suspension,
as illustrated in Figure~\ref{fig:packet_latency}.  In
the figure, the packet that arrives during the first idle sleep
period endures the remainder of the idle period ($t_{buff}$), in
addition to the forwarding time ($t_{proc}$), which includes looking
up relevant header state and translating between the layer 3 \tun
interface and layer 4 NIO sockets. However, the packet that arrives
when polling is active does not experience the idle period overhead.

We now analyze the latency incurred by packets when running
\name. Specifically, we focus on the impact of \ic and \is and the
tradeoff between latency and CPU overhead.
Figure~\ref{fig:latency_overhead} shows the results of our
experiments for UDP.  When \ic=1~cycle the latency closely follows
\is because \name's aggressive sleeping renders it more likely for
packets to arrive when the system is idle, therefore delaying them for up to \is~ms before being processed. With \ic=100~cycles
and \is=100~ms we find about 60~ms of extra delay. Reducing
\is to 10~ms while keeping \ic at 100~cycles reduces latency to as
low as 3.4~ms. We find similar patterns for TCP connections.  In
Figure~\ref{fig:tcp_lat_overhead} we plot connection establishment
times for the TCP echo client and server.  As expected, high values
of \ic and coupled with low \is settings results in quicker
connection establishment. In fact, when the RTT of the link drops
below the time it takes to reach \ic cycles, \name processes all packets in
the TCP handshake without the \fwder going into idle state.

Finally, we consider the latency incurred by a packet during processing and 
forwarding ($t_{proc}$). 
To get a sense of how the latter affects performance, we evaluate $t_{proc}$ 
while running our speed-test app. 
Table~\ref{table:fw_io_perf} shows the results of 
our speed-test for TCP and UDP connections. Processing times for 
established flows are 141~$\mu$s for TCP and 76~$\mu$ s for UDP, indicating 
that the packet forwarding is not a bottleneck for \name's performance. 
The processing times for new connections prove larger, especially for TCP, because of
the overhead of initiating state for the connection.

\begin{table}[t!]
\centering
\scriptsize
\begin{tabular}{|l|c|c|c| }
\hline
 & {\bf Downlink} & {\bf Uplink} & {\bf New Flow} \\ \hline
TCP $t_{proc}$ ($\mu$ s) & 141.6$\pm$0.5 & 275.6$\pm$3.5 & 5,647.8$\pm$998.5  \\ \hline
UDP $t_{proc}$ ($\mu$ s) & 76.6$\pm$2.2 & 230.8$\pm$4.8 & 2,980.8$\pm$224.9 \\ \hline
\end{tabular}
\caption{Mean processing time ($t_{proc}$) and standard error of mean (SEM) 
  for \name's forwarding operations for TCP and UDP flows under stress
  conditions. The first packet of a new flow requires a higher processing time. }
\label{table:fw_io_perf}
\end{table}

\subsection{Throughput of \name} 
\label{sec:intelserveval}

We now  investigate the maximum throughput the system can 
achieve. We use our speed-test app to measure the 
throughput for non-TLS TCP and UDP flows with \is=$\{10~ms,100~ms\}$ and
\ic=100~cycles. This  
setting provides us with a good compromise between CPU usage and latency. 
  
Figure~\ref{fig:throughput_overhead} shows the maximum throughput achieved
by \name's \fwder. We find that \name can provide up to 17.2 and
54.9~Mbps uplink  
and downlink throughput, respectively. 
As expected, when \is increases the throughput decreases, as more packets arrive with
the \fwder in idle state, thus incurring $t_{buffer}$ 
(cf. Figure~\ref{fig:packet_latency}).
\name also has a bias towards downstream traffic, 
which stems from two factors. First, as we discuss
in~\xref{subsec:forwarder}, the NIO read 
operation may potentially 
return multiple packets whereas the \tun 
interface reads only a single packet at a time. Second, the
operations required for  
upstream packets $t_{proc}$ are more computationally expensive 
(see Table~\ref{table:fw_io_perf}). We plan to investigate 
how we can adapt the \cn and \ct parameters to achieve more balanced throughput 
in future work. We note that the performance we report
is still in excess of what is required for modern mobile apps.

Although the \tas operates off-path, the use of thread-safe queues
to enable communication between the \fwder and the \tas and its CPU intensive 
operations can 
inflict significant overhead on traffic throughput. 
As an example analysis task we consider string matching using the
Aho-Corasick algorithm~\cite{aho1975efficient} on the 
traffic to detect tracking. %
Figure~\ref{fig:throughput_overhead} shows \tas's
impact on throughput when performing CPU-intensive string matching on each flow.
In the worst case, \name provides 23.3~Mbps downstream and 10.5~Mbps 
upstream throughput.
Even when stress-testing \name with our speed-test, the maximum queuing time 
endured by packets before the string parsing engine processes them
does not exceed 650~ms. This 
worst-case scenario arises when the queues contain a backlog 
of at least 1,000 packets. Even under such circumstances, the total 
processing time remains low enough 
to provide feedback to users about their traffic in less than a
second (\eg exfiltrated private information).

There remains significant potential for improving the overhead imposed
by the \tas. In particular, we plan to investigate better means of
communicating between the \fwder and the \tas (\eg via Android's IPC~\cite{binderIPC})
to make it more efficient than the thread-safe queue we currently
employ.

\subsection{TLS Performance in \name}
\label{sec:tlsoverhead}

\begin{figure}[t]  
    \centering

    \subfigure[\label{fig:tls_establishment_time}TLS session
	establishment time.]{
      \includegraphics[width=0.92\columnwidth]{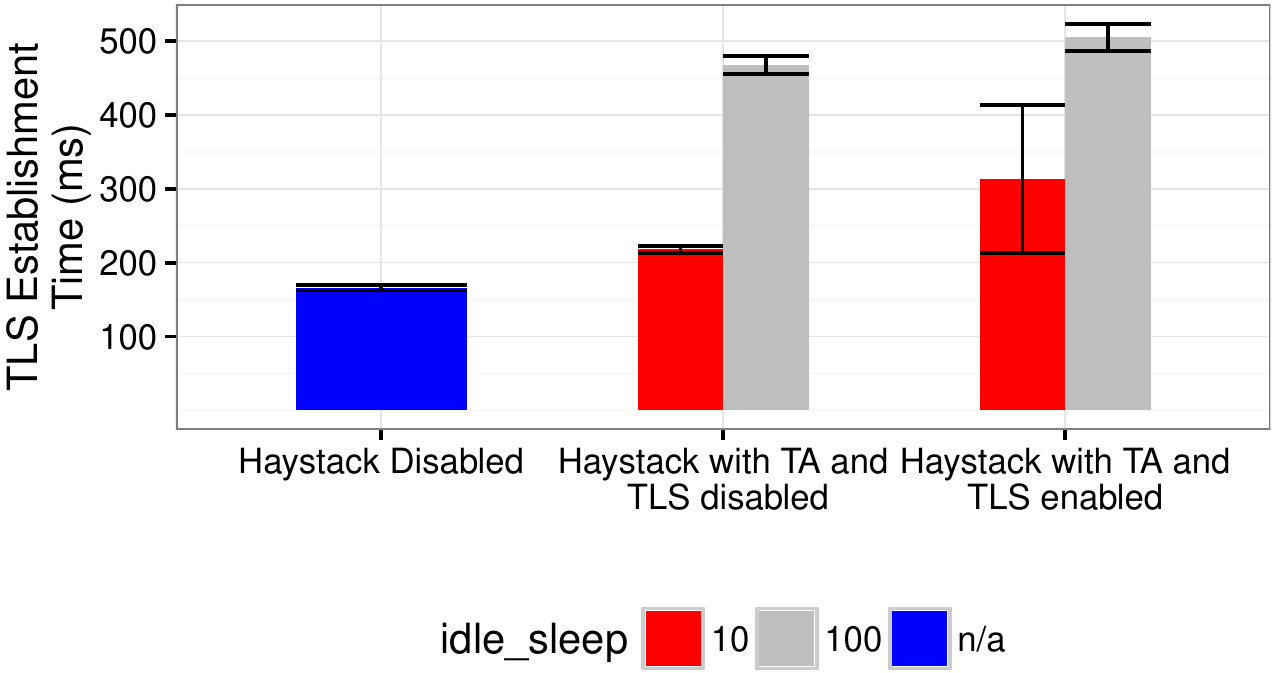}
    }
    \subfigure[\label{fig:tls_downlink_throughput}TLS download
	speeds.]{
      \includegraphics[width=0.92\columnwidth]{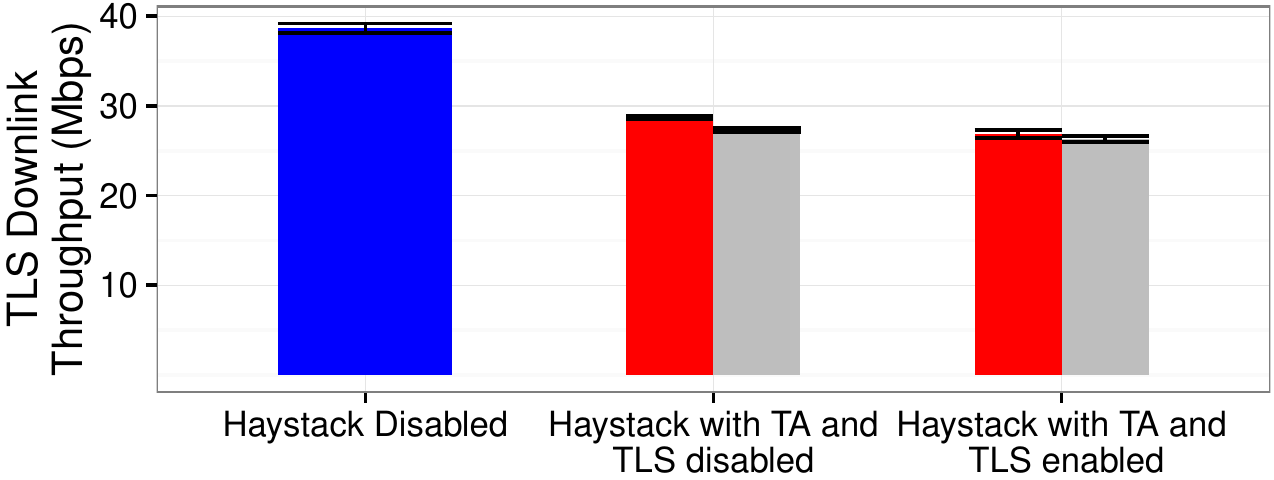}
    }
    \vspace{-1mm}
    \caption{Session establishment and throughput for TLS with \name for 
	different \is configurations, also showing the impact
	of enabling both the TLS proxy and the \tas. We fix \ic to 100 cycles. }
    \label{fig:tls_perf_eval}

\end{figure}

We next turn to the overhead of dealing with encrypted
communication.  Figure~\ref{fig:tls_perf_eval} summarizes the
overhead of the TLS proxy for different configurations.  We first
consider the baseline overhead of \name without the TLS proxy
enabled on TLS connection establishment times, as shown in
Figure~\ref{fig:tls_establishment_time}. With an \is of 10~ms the
  TLS connection establishment time is 218~ms. Increasing \is
to 100~ms has a large effect, doubling the TLS establishment time
(466~ms).  Using the TLS proxy further increases establishment time to 653~ms with \is at 10~ms, and 503~ms with \is at 100~ms.

We next assess the overhead of the TLS proxy on throughput, as shown
in Figure~\ref{fig:tls_downlink_throughput}. Compared to not running
\name at all, the overhead of the TLS proxy is 26\% and 29\% for \is
= 10~ms and 100~ms, respectively.  Despite the decrease in
throughput, overall throughput with the TLS proxy is still 26~Mbps,
which (as discussed in~\xref{sec:cpubatteryeval}) allows
playing a 1080p YouTube video without affecting the user experience.
We find \is has little impact on throughput since subsequent packets
bring the \fwder out of the idle state, thus avoiding $t_{buffer}$
for the bulk of the transfer. The fact that the TLS proxy
reassembles the streams for the \is also helps reduce the overhead.

\subsection{Using \name to Measure Performance}
\label{sec:dns_measurement}

\name's \fwder parameters 
can affect \name's ability to accurately measure network performance. 
This section compares \name's ability to assess traffic and network
performance with {\tt tcpdump} packet-level timestamps on a rooted phone. 
For these experiments, we instrument a rooted mobile device with an 
Android app that performs 500 UDP-based DNS queries to 
8.8.8.8 for {\tt [nonce].stonybrook.edu}. 
The nonce ensures that all queries bypass any intermediate cache.
We perform the DNS lookups in two different settings: 
($i$) when the DNS traffic goes directly through the default
gateway, and ($ii$) when \name forwards the DNS traffic. 
This allows us to calibrate \name by comparing
actual performance as seen by user-space apps with 
passive measurements as seen by \name.

\begin{figure}[t]  
    \centering
    \includegraphics[width=1\columnwidth]{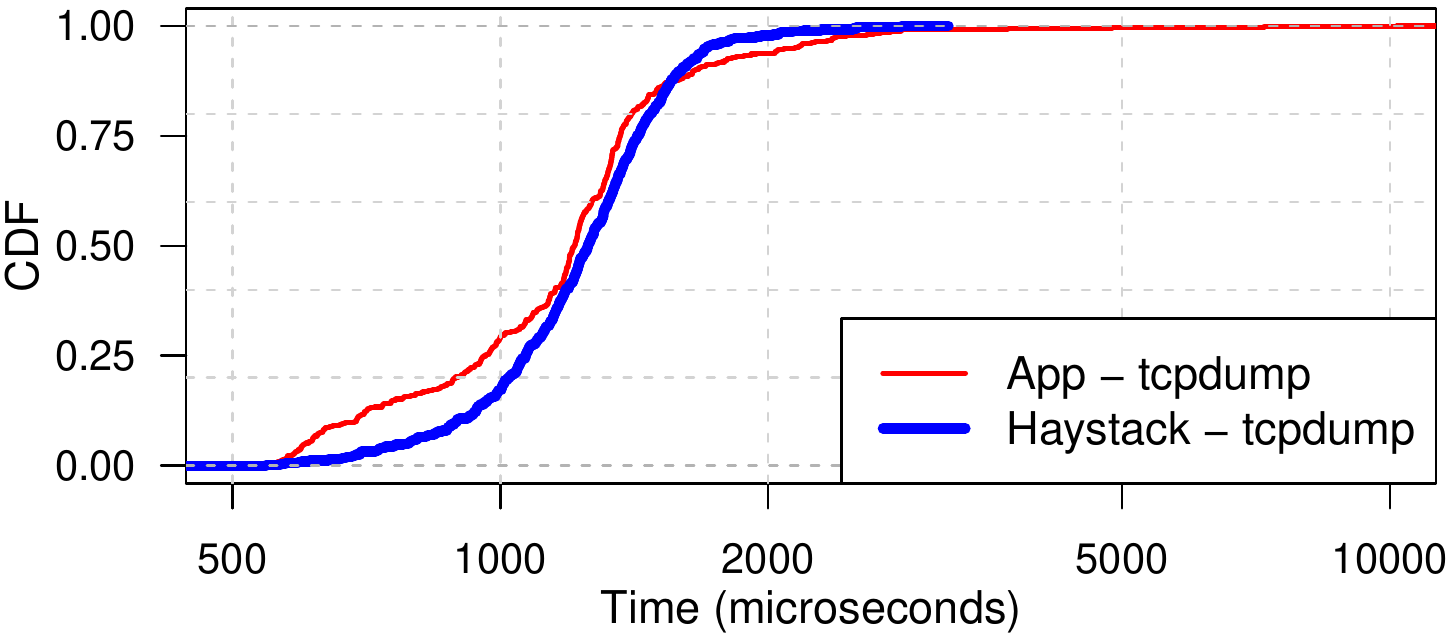}
    \caption{Difference between DNS lookup times as measured by a Java-based 
	application (red line) and by \name (blue line), both compared to {\tt tcpdump}.  The cross between the red line and the blue line is likely due to instabilities in measuring from the application that is introduced by the Java VM on Android.
	The analysis confirms \name as a valid user space performance 
	measurement platform.}
    \label{fig:dns_diffplot}
\end{figure}

We use \is=0~ms and \ic=200~cycles so that we can minimize packet wait time and to 
prevent blocking on a given interface at the expenses of increasing
the CPU load. 
We send the queries sequentially and with random inter-query delays of $250~ms + rand(0,400)~ms$,
over a stable, well-provisioned WiFi link. The random delay ensures that
packets are not queued and that we are sampling the times in different polling states of \name (recall Figure~\ref{fig:packet_latency}).
We factor out transient effects in the network by computing the difference
between measurements made by \name, those made by the Android app,
and those obtained via {\tt tcpdump}. 
Figure~\ref{fig:dns_diffplot} shows the difference between our
user-level measurements and the reference \texttt{tcpdump}
measurements.   
Table~\ref{table:dns_timing_stats} summarizes these differences.
The difference between 
\name's observation of DNS latency and the Android app is
small, with 
mean and median values differing by less than 50~$\mu$s. 
We find similar results over a cellular link, which we expect because
the measured overheads stem from the Java virtual machine and 
\name, not from varying
network conditions.  The magnitude of the differences we observe remains
orders of magnitude smaller than typical mobile network delays,
making \name suitable for fine-grained network performance measurements.

\begin{table}[t!]
\centering
\scriptsize
\begin{tabular}{|c|c|c|c| } %
\hline
\bf  & \bf Mean & \bf Median & \bf StDev \\ %
\hline
\name-tcpdump & $1,261 \mu$s & $1,254 \mu$s & $303 \mu$s \\ %
\hline
App-tcpdump & $1,250 \mu$s & $1,211 \mu$s & $658 \mu$s \\ %
\hline
\end{tabular}
\caption{Detail statistics of the distribution shown in 
  Figure~\ref{fig:dns_diffplot}.}
\label{table:dns_timing_stats}
\end{table}

\subsection{\name Adaptability}

Above we demonstrate the tradeoffs between resource usage and
performance of \name as controlled by varying \ic and \is parameters.  While
we find no sweet spot that is ideal in all circumstances, our
observations point to three possible ways to adapt \name's operation
to the current device state.

\custompara{Adapting parameters: } Our first technique for reducing
\name's overhead is adapting the parameters to different usage
scenarios to strike a balance for the current device
state.  We consider two scenarios: ($i$) when the user interacts
with the device or in the presence of latency-sensitive background traffic
(\eg streaming audio), and ($ii$) when the phone is idle with minimal
network activity (\eg push notifications), a critical scenario 
as phones remain idle the 
majority of time~\cite{aucinas2013staying,vallina2012energy}. In the first case,
minimizing latency and guaranteeing the best user experience is
critical (``performance'' mode), whereas in the second case, traffic
is delay tolerant (``low-power'' mode)~\cite{aucinas2013staying}.
Based on our results we determine that \is = 10~ms and \ic = 100~cycles
gives the best tradeoff of performance and resource usage for
delay-sensitive usage and \is = 100~ms and \ic = 100~cycles gives the
best tradeoff during delay-insensitive usage.
Table~\ref{table:hs_conf} summarizes the overheads and performance
for each of these settings.
 
\custompara{Sampling:} A second way to reduce the overhead of \name
is to sample a fraction of the connections or application sessions
to fully analyze.  This has the usual costs and benefits of
sampling: lower resource usage (\eg by not requiring the \tas to do
as much work) on the one hand and less coverage on the other hand.
We believe that users likely care most about the apps and networks
they use regularly and therefore while sampling may take longer to
uncover issues, these issues will in fact be uncovered due to high
usage. 

\custompara{Targeting:} A final technique to reduce the overhead of
\name is to allow users to instruct \name to only consider certain
apps.  For instance, this may be useful when the user installs a new
app and wants to understand what information it may leak.

\begin{table*}[t!]
\scriptsize
\centering
\begin{tabular}{|l|c|c|C{1.58cm}|C{1.8cm}|C{2.1cm}|C{2.68cm}|C{1.3cm}|}
\hline
{\bf Mode} & {\bf idle\_sleep } & {\bf max\_idle\_cycles } &  {\bf Mean UDP RTT (ms)} & {\bf Mean TCP Conn. Time (ms)} & {\bf Mean 
	TLS Conn. Time (ms)} & {\bf Max. Throughput [Up/Down] (Mbps)} 
	& {\bf Mean CPU (\%)} \\ \hline
Performance & 10 & 100 & 5.4 & 24.8 & 313.1 & 17.2/54.9 & 11.2 \\ \hline
Low-power & 100 & 100 & 60.8 & 65.3 & 505.3 & 16.7/48.2 & 2.7 \\ \hline
\end{tabular}
\vspace{-1mm}
\caption{Summary of \name's performance for each 
  operational mode in a 5~GHz WiFi link with 3~ms RTT. }
\label{table:hs_conf}
\end{table*}

\begin{table}[t!]
\scriptsize
\centering
\begin{tabular}{|c|c|c|c|}
\hline
 {\bf Users} & {\bf Apps} & {\bf Total Flows} & {\bf Domains} \\
\hline
450 & \trackedapps & \trackedflows & \uniquedomains \\
\hline
\end{tabular}
\vspace{-1mm}
\caption{Summary and scale of our user study.}
\label{table:user_study}
\end{table}

\section{Advantages of Haystack}
\label{sec:applications}

As we illustrate in \xref{sec:background}, the research community
has focused much attention on understanding mobile devices and
networks.  These previous efforts have produced many significant
insights.  We now turn to discussing \name's advantages---for both
researchers and users---relative to the state of the art in mobile
measurement and app profiling.  We describe \name capabilities, as well as early
results culled from data collected by the 450~users who have
installed \name to date.  We summarize our initial dataset in
Table~\ref{table:user_study}.  We stress that these are not
full-fledged measurement studies and are presented for illustrative
purposes.  Finally, we note that while we discuss \name's
capabilities in isolation they can often be used together to an even
greater effect.

\parax{Unprecedented View of Encrypted Traffic}
\name's TLS proxy allows analysis---with the user's permission---of
encrypted communication.  This information remains opaque to other
methodologies (e.g., ISP network traces) or requires trusting a
third-party middleman to decrypt and correctly re-encrypt traffic
while protecting the clear-text version (e.g., remote VPN
endpoints).  This visibility is significant as we find that in our
dataset 22\% of the flows are encrypted and less than 20\% of apps
send all their traffic in the clear.  Therefore, gaining an
understanding of the information flow within the mobile ecosystem
critically depends on being able to cope with encrypted traffic.

\parax{Unprecedented View of Local Traffic}
\name can naturally observe local network traffic that never
traverses the wide-area network.  This traffic does not appear in ISP
network traces \cite{Narseo:IMC2012,Gill:2013:BPF:2504730.2504768}
and methodologies that rely on remote VPN tunnels \cite{meddle}.
This capability will only increase in importance given the emergence
of Internet-Of-Things (IoT) devices in the household, typically using
mobile devices for control.
Our dataset includes 40~apps that generate local traffic,
ranging from baby monitors to media servers to smart TV remote
control apps.

\parax{Representative View of Apps}
When trying to understand app behavior a natural first question concerns
finding a set of apps to study.  \name answers this question quite
simply by considering the natural set of apps each user executes.
For other methodologies---such as static and dynamic analysis---the
set of apps considered often results from a crawl of the Google
Play store.  However, this neglects built-in apps and apps from
non-standard or alternative app stores \cite{enck2010taintdroid},
as well as behavioral changes caused by new app updates. 
Further, such studies frequently exclude non-free apps or code paths that
stem from an in-app purchase \cite{fahl2012eve,enck2010taintdroid}.
\name includes all of these aspects naturally.

Our initial data suggests analyzing these apps is important.  We
find 15\% of the apps we observe did not come from the Google Play
store and include ($i$) apps developed by large and small device
vendors and mobile carriers, ($ii$) pre-installed third-party apps
(e.g., Kineto, a Wifi calling app~\cite{kineto}) and ($iii$) apps
downloaded from alternate or regional app stores~\cite{petsas2013rise}.  
These apps create 22\% of the traffic we observe.  Further, we find
that 3.7\% of the apps in Google Play are not free and 29\% of the
apps include in-app purchases.  Both of these represent code paths
often skipped when studying app behavior, but which
\name considers as a matter of course.  Finally, we find that apps
not originating from the Google Play store do in fact leak personal
information and unique identifiers (\eg to third-party tracking
services such as Crashlytics).

\parax{Representative Code Paths}
Related to the last point is that \name naturally deals with common
and esoteric code paths.  Dynamic analysis requires manual
navigation within apps, oftentimes synthetically generated
via UI Monkeys~\cite{wong2016intellidroid}.  
The results in turn prove sensitive to
details of this test navigation.  \name does not suffer from this
problem because the interactions observed reflect the natural way that
the user interacts with the app.  Therefore, even though \name will
never know that some unused code path is nefarious, it does not
matter to that user because that user never invokes the problematic
case.  On the contrary, a dynamic analysis test case might miss some
buried feature that gets exercised by real users---in
which case \name will catch networking activity resulting from that code path.

\parax{User Involvement}
\name's position on user devices provides the capability to engage
users in novel ways that benefit both research into the mobile
ecosystem, but also the users themselves.  For instance, users could
be given the option to opt-in to assisting researchers assessing
Quality-of-Experience (QoE) of their normal traffic.  Combining
qualitative user feedback with quantitative measurements of the
traffic could be a powerful combination that brings many insights to
this space.  Another case where direct interaction helps the user is
in understanding the leakage of personal information from the
phone.  By exposing this information to users they can make more
informed decisions about what apps to use or what permissions to
grant to specific apps.

\parax{Novel Policy Enforcement} 
\name's position between apps and the network provides for a unique
ability to implement policy before traffic leaves the mobile device.
While \name can certainly enforce traditional network
policies---\eg blocking access to specific IP addresses or hostnames, or
rate-limiting certain traffic---the insight into app content means
the policies can be more semantically rich (\eg blocking based on
attempted leaking of a specific piece of information like the
IMEI to an untrusted online service).  
Further, policies can be richer than simple blocking
decisions.  For instance, specific traffic could be sent through a
VPN tunnel or an anonymization network.  Or, particular sensitive
elements of the contents could be obfuscated (\eg providing a
random IMEI to each app to hinder cross-app tracking).

\parax{Enabling Reactive Measurement}
\name is not beholden to naturally occurring traffic, but can
trigger its own active measurements as needed.  These
measurements---taken from the device's natural perspective---could
be proactive in an attempt to better understand the current network
context.  Alternatively, active measurements could be taken
reactively \cite{allman2008reactive} based on some
observation---\eg to diagnose slow transfers or delay spikes.
Finally, active measurements could explore ``what if'' sorts of
analysis that could in turn be used to tune the device's operation
for better performance.  For instance, \name could explore whether
an alternate DNS resolver would yield faster or better answers
compared with the standard resolver.

\parax{Explicitness Leading To Precision}
Many of the problems in measuring the mobile ecosystem stem from the
need to infer or estimate various aspects of the communication.
\name gets around many of these problems because it operates within
explicit context from the mobile operating system.  For instance,
instead of inferring a device's phone number, \name has a direct
understanding of the value and therefore can directly hunt for it in
the traffic instead of searching for something that ``looks like''
a phone number and then trying to determine if that is in fact the
device's phone number or not.

\begin{figure}[t!]  \centering
  \includegraphics[width=1\columnwidth]{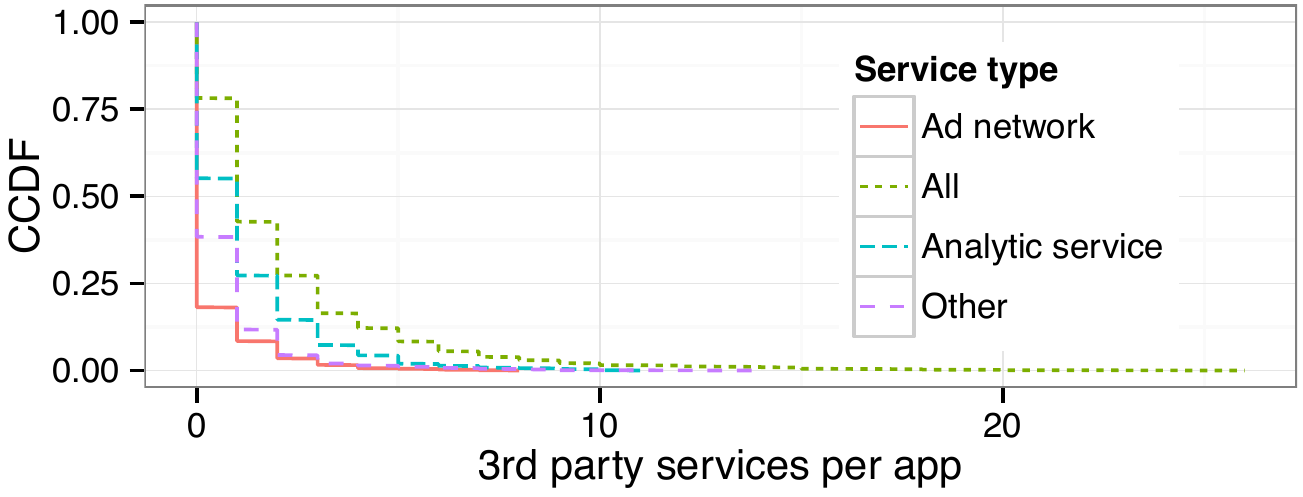}  
  \caption{Distribution of the number of third-party services per
    app across our dataset.}
  \label{fig:ccdf_thirdpartiesperapp}
\end{figure}

As another example, \name is able to directly attribute traffic
flows to apps rather than using heuristics or inferences.  Within
our dataset we use this ability to detect apps that use third-party
services---for myriad reasons, including ad delivery, analytics,
alternative push notifications \cite{jpush}---at the network level.
While a simple app-agnostic count of accesses to specific services
provides some understanding of popular services, this method leaves
ambiguous whether the popularity stems from broad use across apps or
simply use by popular apps.  \name can directly answer this
question.  Figure~\ref{fig:ccdf_thirdpartiesperapp} shows the
distribution of the number of third-party services used per app
across our dataset.  By ranking the online services by the number of
apps connecting to them, we can see that Crashlytics and Flurry are
the most common third-party services across our corpus of mobile
apps.

\parax{Large-Scale Deployment}
\name's operation as a normal user-level app that does not require
rooting a device or a custom firmware version means the barrier to
entry for using \name is low.  This is useful for research purposes
as we can coax more users to the platform than a more cumbersome
tool would require.  Further, the platform gives us a direct path to
moving beyond research and to actually helping normal users
understand the operation of their devices that simply would not be
possible if users had to jump through significant hoops to install
and use \name.  These two aspects feed a virtuous circle: more users
provide more data that we can leverage to increase users'
understanding, which in turn provides a larger incentive to entice
additional users.

While each of the above are advantages in their own right, they
become even more powerful when combined.  For instance, \name has
the power to understand that a specific app is trying to leak their
phone number to some IoT device in their house within an encrypted
connection.  Further, this understanding could be communicated to
the user, as well as serving as fodder for a policy that thwarts
such activity in the future.  Each aspect of this example is either
impossible with current techniques or at least requires inference
and heuristics.

\section{Discussion and Future Work}

In~\xref{sec:applications} we have discussed how \name's features
provide an unprecedented window into the mobile ecosystem.  \name
shares traits of other network monitoring/measurement platforms we
maintain~\cite{broweb,netalyzr:imc}, increasing our confidence that
\name's architecture provides a solid basis for future
exploration. Since many enterprise and mobile device management
solutions rely on the VPN permission, we also expect that Android will
continue to support it.

While \name's implementation runs on the Android, support on other
platforms proves feasible: recent iOS releases offer underlying API
primitives similar to those enabling \name on Android~\cite{iosVPN}.
In fact, given iOS's tighter technical constraints to app development
(\eg preventing Taintdroid-like approaches), \name's approach provides
a promising avenue for investigating the iOS ecosystem.

We stress that the applications of \name sketched in this paper serve
to exemplify its abilities as a traffic inspection platform, not as a
step in the network security arms race.  For example, we do
\textit{not} advocate \name as a full-blown TLS inspector---rather, we
demonstrate that the platform supports development in this direction
to a point that can readily provide interesting results.

We are currently exploring ways to open up \name to the research and
app developer communities.
By further separating the Forwarder and Traffic Analyzer components we
can establish access to the device's traffic streams for other apps,
effectively providing a proxy to the absent ``packet capture''
permission on Android.\footnote{Personal communication with the Google
  Android team suggests that this option remains unlikely to ever
  materialize directly in the Android OS.}
In doing so, we can overcome an additional constraint of Android's
security model, namely that only a single VPN app can run at any given
time.

We acknowledge that opening up \name's capabilities to third-party
apps would raise grave security concerns, as malicious apps may abuse
\name's capabilities for nefarious purposes. We defer full treatment
to future work and here only mention that Android's custom permissions
model~\cite{custompermissions} provides avenues for making access to
these new capabilities controllable by the user.

We are planning to open-source the \name codebase, and will make
anonymized data collected by \name available online via a web-based
query interface similar to ReCon~\cite{recon} or
Censys.io~\cite{censys}.

\section{Summary}
\label{sec:summary}

We have presented the design, implementation, and  evaluation
of \name, a multi-purpose mobile vantage point
for Android devices built on top of Android's VPN permission. 
As \name runs completely in user-space, it enables large-scale measurements 
of real-world mobile network traffic from end-user devices, with organic 
user and network input.

Through extensive evaluation, we have demonstrated that \name realizes
a flexible mobile measurement
platform that can deliver sufficient
performance with modest resource overhead and minimal impact on
user activity when compared to state-of-the-art methods
that rely on static and dynamic analysis. 

\name opens a new horizon in mobile research by 
achieving an architectural sweet-spot that makes it easy to
install on regular user phones (thus enabling large-scale deployment
and benefiting from user's input)
while enabling in-depth visibility into device activity and
traffic (thus providing installation incentives to the user).
Using a deployment to 450 users who installed the \name app from Google's Play Store, we
demonstrated \name's ability to provide meaningful insights about 
protocol usage, its ability to identify security and privacy
concerns of mobile apps, and to characterize mobile traffic performance.

\end{sloppypar}

\newpage

{\footnotesize \bibliographystyle{acm}
\bibliography{paper.bib}}

\end{document}